\documentstyle[aps]{revtex}

\begin{document}
\author{Sheng Li\thanks{%
lisheng@itp.ac.cn}, Yong Zhang\thanks{%
zhangyo@itp.ac.cn} and Zhongyuan Zhu\thanks{%
zzy@itp.ac.cn}}
\title{Topological Defects in 3-d Euclidean Gravity}
\address{Institute of Theoretical Physics, Academia Sinica, PO Box 2735, Beijing
100080, P.R. China}
\draft
\maketitle

\begin{abstract}
By making use of the complete decomposition of $SO(3)$ spin connection, the
topological defect in 3-dimensional Euclidean gravity is studied in detail.
The topological structure of disclination is given as the combination of a
monopole structure and a vortex structure. Furthermore, the Kac-Moody
algebra generated by the monopole and vortex is discussed in three
dimensional Chern-Simons theory.
\end{abstract}


\section{Introduction}

An exciting development in cosmology has been the realization that the
universe may behave very much like a condensed matter system. Analogous to
those found in some condensed matter systems$-$vertex line in liquid helium,
flux tubes in type-II superconductors, or disclination lines in liquid
crystals, the topological space-time defects may have been found at phase
transitions in the early history of the universe. They may help to explain
some of the largest-scale structure seen in the universe today.

As a kind of topological defect, the disclination is caused by inserting
solid angles into the flat space-time. In Riemann-Cartan geometry, this
effect is showed by the integral of the affine curvature along a closed
surface. Duan, Duan and Zhang\cite{Duan-Duan} had discussed the
disclinations in deformable material media by applying the gauge field
theory and decomposition theory of gauge potential. In their works, the
projection of disclination density along the gauge parallel vector was found
corresponding to a set of isolated disclinations in the three dimensional
sense and being topologically quantized. Furthermore, the space-time
disclinations in 4-dimensional with Euclidean signature and Lorentz
signature were discussed by Duan and Li\cite{duan-li,li}, similar results
were obtained..

In this paper, we discuss the disclinations in three dimensional gravity by
making use of the decomposition formula of $SO(3)$ gauge potential proposed
by Faddeev and Niemi\cite{faddeev}. This decomposition theory provides new
tool to study the topological defects. In their decomposition, a complex
field $\phi $ is introduced naturally from gauge transformation. By studying
the transformation properties, they showed this complex field $\phi $ and
the projection of the $SU(2)$ gauge potential of a unit vector $n^a$ form a
multiplet. Just like the magnetic monopole theory, by introducing the
Abelian projection of $SO(3)$ gauge field, we define the topological charge,
which is combined by monopoles and vortices. We find it is the complex field 
$\phi $ whose zero points behave as the sources of the vortices. The
projection of space-time disclination density along the gauge parallel
vector is topological quantized with $2\pi $ as the unit solid angle.
Further, by making using of the three dimensional Chern-Simons gravity
theory (see for examples\cite{bal,brown,ba4,ba1,ba2,ba3}), we show the
monopole charges and the vorticities can server as the generator of
Kac-Moody algebra.

This paper is arranged as follows: In section II, we introduce the
definitions of the topological defects in 3-dimensional gravity. In section
III, by using the decomposition of $SO(3)$ gauge potential, we discuss the
relationship of disclination and topological charge. The local structure of
the topological defect is given in section IV. At last, we discuss the
algebra generated by monopole and vortex in three dimensional Chern-Simons
theory in section V.

\section{Topological defects in 3-dimensional Euclidean gravity}

As it was shown in \cite{Duan-Duan}, the dislocation and disclination
continuum can be described by the reference, deformed and natural states.
For the natural state there is only an anholonomic rectangular coordinate $%
Z^a$ ($a=1,2,3$) and 
\begin{equation}
\delta Z^a=e_\mu ^adx^\mu ,
\end{equation}
where $e_\mu ^a$ is triad. The metric tensor of the Riemann-Cartan manifold
of natural state is defined by 
\begin{equation}
g_{\mu \nu }=e_\mu ^ae_\nu ^a.
\end{equation}

We have known that the metric tensor $g_{\mu \nu }$ is invariant under the
local $SO(3)$ transformation of triad. The corresponding gauge covariant
derivative 1-form of a vector field $\phi ^a$ on ${\bf M}$ is given as 
\begin{equation}
D\phi ^a=d\phi ^a-\omega ^{ab}\phi ^c,
\end{equation}
where $\omega ^{ab}$ is $SO(3)$ spin connection 1-form 
\begin{equation}
\omega ^{ab}=-\omega ^{ba}\quad \quad \omega ^{ab}=\omega _\mu ^{ab}dx^\mu .
\end{equation}
The affine connection of the Riemann--Cartan space is determined by 
\begin{equation}
\Gamma _{\mu \nu }^\lambda =e^{\lambda a}D_\mu e_\nu ^a.
\end{equation}
The torsion tensor is the antisymmetric part of $\Gamma _{\mu \nu }$ and is
expressed as 
\begin{equation}
T_{\mu \nu }^\lambda =e^{\lambda a}T_{\mu \nu }^a,
\end{equation}
where 
\begin{equation}
T_{\mu \nu }^a=\frac 12(D_\mu e_\nu ^a-D_\nu e_\mu ^a).
\end{equation}
The Riemannian curvature tensor is equivalent to the $SO(3)$ gauge field
strength tensor 2-form $F^{ab}$, which is given by 
\begin{equation}
F^{ab}=d\omega ^{ab}-\omega ^{ac}\wedge \omega ^{cb}\quad \quad F^{ab}=\frac %
12F_{\mu \nu }^{ab}dx^\mu \wedge dx^\nu
\end{equation}
and relates with the Riemann curvature tensor by 
\begin{equation}
F_{\mu \nu }^{ab}=-R_{\mu \nu \sigma }^\lambda e_\lambda ^ae^{\sigma b}.
\end{equation}

The dislocation density is defined by 
\begin{equation}
\alpha ^a=T^a\quad \quad \quad T^a=\frac 12T_{\mu \nu }^adx^\mu \wedge
dx^\nu .
\end{equation}
Analogous to the definition of the 3-dimensional disclination density in the
gauge field theory of condensed matter, we define the space-time
disclination density as 
\begin{equation}
\theta ^a=\frac 12\varepsilon ^{abc}R_{\mu \nu \sigma }^\lambda e_\lambda
^be^{\sigma c}dx^\mu \wedge dx^\nu =-\varepsilon ^{abc}F^{bc}.
\label{defination-disclination}
\end{equation}

The size of the space-time disclination can be represented by the means of
the surface integral of the projection of the space-time disclination
density along an unit vector field $n^a$ 
\begin{equation}
S=\oint_\Sigma \theta ^an^a=-\oint_\Sigma \varepsilon ^{abc}n^aF^{bc}
\label{4w}
\end{equation}
where $\Sigma \,$ is a closed surface including the disclinations. The new
quantity $S$ defined by (\ref{4w}) is dimensionless. Using the so-called $%
\phi $-mapping method and the decomposition of gauge potential Y. S. Duan et
al have proved that the dislocation flux is quantized in units of the Planck
length\cite{Duan4} and the disclination in condensed matter is quantized
also in similar structure as that of magnetic monopole. In this paper, using
the new decomposition formula given by Faddeev and Niemi\cite{faddeev}, we
will show that apart from the monopole structure, a vortex structure also
contributes to the disclination.

\section{Decomposition of $SO(3)$ gauge potential and topological charge of
Abelian projection of $SO(3)$ gauge field}

For $so(3)$ Lie algebra is homomorphic to $su(2)$ Lie algebra, we can
introduce the $SU(2)$ gauge potential in term of $SO(3)$ spin connection by 
\begin{equation}
\omega ^a=\frac 12\varepsilon ^{abc}\omega ^{bc}.
\end{equation}
The corresponding gauge field 2-form is 
\begin{equation}
F^a=\frac 12\varepsilon ^{abc}F^{bc}=dA^a+\frac 12\varepsilon
^{abc}A^b\wedge A^c.
\end{equation}
The covariant derivative of $n^a$ is 
\begin{equation}
Dn^a=dn^a-\varepsilon ^{abc}\omega ^bn^c.
\end{equation}
From this equation we can solve the gauge potential 1-form $\omega ^a$
expressed in term of $n^a$ as 
\begin{equation}
\omega ^a=An^a+\varepsilon ^{abc}dn^bn^c-\varepsilon ^{abc}Dn^bn^c,
\end{equation}
and curvature 2-form is 
\begin{equation}
F^a=-\frac 12\varepsilon ^{abc}dn^b\wedge dn^c+\frac 12\varepsilon
^{abc}Dn^b\wedge Dn^c+\varepsilon ^{abc}n^bdDn^c+Dn^a\wedge A+n^adA.
\end{equation}
Recently Faddeev and Niemi\cite{faddeev} showed the covariant part of $%
\omega ^a$ can be expressed as 
\begin{equation}
-\varepsilon ^{abc}Dn^bn^c=\rho dn^a+\sigma \varepsilon ^{abc}dn^bn^c
\end{equation}
where $\rho $ and $\sigma $ are coefficients and can be combined into a
complex field 
\begin{equation}
\phi =\rho +i\sigma .
\end{equation}
Then the garge potential 1-form is rewritten as 
\begin{equation}
\omega ^a=An^a+\varepsilon ^{abc}dn^bn^c+\rho dn^a+\sigma \varepsilon
^{abc}dn^bn^c.
\end{equation}
Under a $SU(2)$ gauge transformation generated by $\alpha ^a=\alpha n^a$ the
functional form of above equation remain intact and $\phi $ transforms as a $%
U(1)$ vector field 
\begin{equation}
\phi \rightarrow e^{i\alpha }\phi 
\end{equation}
and $A$ transforms as 
\[
A\rightarrow A+d\alpha ,
\]
which means the multiplet $(A_\mu ,\phi )$ transforms like the field
multiplet in abelian Higgs model. We can introduce the $U(1)$ covariant
derivative of $\phi $%
\begin{eqnarray}
D\phi  &=&d\phi -iA\phi   \nonumber \\
&=&d\rho +A\sigma +i(d\sigma -A\rho )  \nonumber \\
&=&D\rho +iD\sigma .  \label{dphi}
\end{eqnarray}
The curvature 2-form can be rewritten as 
\begin{equation}
F^a=n^a(dA-\frac 12(1-||\phi ||^2)\varepsilon ^{abc}dn^b\wedge dn^c+D\rho
\wedge dn^a+\varepsilon ^{abc}D\sigma \wedge dn^bn^c.  \label{f-phi}
\end{equation}
Solving from the equation (\ref{dphi}), we get the decomposition of $A$ as 
\begin{equation}
A=\frac 1{2i||\phi ||^2}(d\phi \phi ^{*}-\phi d\phi ^{*}-D\phi \phi
^{*}+\phi D\phi ^{*}).  \label{a}
\end{equation}
By making using of equation (\ref{f-phi}), we get the disclination density
projection as 
\begin{equation}
F^an^a=f-(1-||\phi ||^2)K,  \label{b1}
\end{equation}
where $f$ is $U(1)$-like curvature corresponding to the projection of $SO(3)$
spin connection 
\begin{equation}
f=dA
\end{equation}
and $K$ is the solid angle density 
\begin{equation}
K=\frac 12\varepsilon ^{abc}n^adn^b\wedge dn^c=d\Omega .  \label{b2}
\end{equation}
Using (\ref{b1})-(\ref{b2}), we obtain the surface integral (\ref{4w}) as 
\begin{equation}
S=-\int_\Sigma f+\int_\Sigma (1-||\phi ||^2)K.  \label{d-c}
\end{equation}
The topological charge of the Abelian projection is defined as a surface
integral of induced $U(1)$-like gauge field over a closed surface $\Sigma $ 
\begin{equation}
Q=\frac 1{4\pi }\int_\Sigma F
\end{equation}
where $F$ is the topological charge density defined by 
\begin{eqnarray}
F &=&-F^an^a+\frac 12\varepsilon ^{abc}n^aDn^b\wedge Dn^c  \nonumber \\
&=&\frac 12\varepsilon ^{abc}n^adn^b\wedge dn^c-dA.  \label{t-c-d}
\end{eqnarray}
The first term of the right hand of (\ref{t-c-d}) is a 3-dimensional solid
angle density which present a monopole structure and the second term
presents a vortex structure. The topological charge is then expressed as 
\begin{equation}
Q=\frac 1{4\pi }\int_\Sigma K-\frac 1{4\pi }\int_\Sigma f,  \label{t-c}
\end{equation}
which means the topological charge equals to the sum of the solid angle of
the closed surface $\Sigma $ (or equatively the monopole charge) and half of
the vorticity corresponding to the Abelian projection of $SO(3)$ spin
connection. This topological characteristic labels the topological property
of the $SO(3)$ gauge field by fixing the coset $SO(3)/U(1).$

Comparing equation (\ref{d-c}) with equation (\ref{t-c}) we see the term $%
\phi K$ distinguish different geometries from the same topology. When $%
||\phi ||$ is taken as a constant the disclination is topological quantized. 
\[
S=(1-||\phi ||^2)\int_\Sigma K-\int_\Sigma f.
\]
Moreover, if we choose the unit vector $n^a$ as a gauge constant 
\[
Dn^a=0,
\]
we find the disclination is just the topological charge we defined, or in
other words, the projection of the disclination on a gauge constant is
topological quantized 
\[
S=\int_\Sigma K-\int_\Sigma f=4\pi Q.
\]
On the other hand, for $A_\mu $ and $\phi $ form a multiplet, the
topological charge and topological structure of the vortex corresponding to $%
A_\mu $ can be expressed directly by the complex field $\phi $ as what will
be shown at the following section.

\section{Local topological structure of the topological defects}

In this section, we will study in detail the local topological structure of
the topological defects.

By making use of the stokes theorem, it seems the topological charge $Q$ is
equals to zero for $K$ and $f$ are exact 
\begin{eqnarray}
dK &=&d^2\Omega =0;  \label{s-a} \\
df &=&d^2A=0.  \label{s-b}
\end{eqnarray}
However, it is not so when considering the singularity of the fields. It
will be shown in the follows that the equation (\ref{s-a}) and (\ref{s-b})
are satisfied ``almost everywhere'' except some singular points. Defects
introduce source terms at the right-hand sides for defects present points
where derivatives fail to commute.

The unit vector $n^a$ can be expressed as

\begin{equation}
n^a=\frac{\varphi ^a}{||\varphi ||}
\end{equation}
where $\varphi ^a$ is the vector along the direction of $n^a$. There exists
the relations 
\begin{equation}
dn^a=\frac 1{||\varphi ||}d\varphi ^a+\varphi ^ad\frac 1{||\varphi ||}
\end{equation}
and 
\begin{equation}
\frac \partial {\partial \varphi ^a}\frac 1{||\varphi ||}=-\frac{\varphi ^a}{%
||\varphi ||^3}.
\end{equation}
By making use of above formulas we find 
\begin{eqnarray}
d^2\Omega  &=&\frac 12\varepsilon ^{abc}dn^a\wedge dn^b\wedge dn^c  \nonumber
\\
&=&\frac 12\varepsilon ^{abc}d(n^adn^b\wedge dn^c)  \nonumber \\
&=&\frac 12\varepsilon ^{abc}d(\frac{\varphi ^a}{||\varphi ||^3})\wedge
d\varphi ^b\wedge d\varphi ^c  \nonumber \\
&=&-\frac 12\varepsilon ^{abc}(\frac \partial {\partial \varphi ^a}\frac %
\partial {\partial \varphi ^d}\frac 1{||\varphi ||^3})d\varphi ^d\wedge
d\varphi ^b\wedge d\varphi ^c.
\end{eqnarray}
Define the Jacobian $J(\frac \varphi x)$%
\begin{equation}
\varepsilon ^{abc}J(\frac \varphi x)=\varepsilon ^{\mu \nu \lambda }\partial
_\mu \varphi ^a\partial _\nu \varphi ^b\partial _\lambda \varphi ^c
\end{equation}
and make use of the Laplacian relation 
\begin{equation}
\frac \partial {\partial \varphi ^a}\frac \partial {\partial \varphi ^a}%
\frac 1{||\varphi ||^3}=-4\pi \delta ^3(\varphi ).
\end{equation}
Then finally we get 
\begin{equation}
d^2\Omega =4\pi \delta ^3(\varphi )J(\frac \varphi x)d^3x.  \label{ddw}
\end{equation}
It means that the solid angle density is exact everywhere except the zeroes
of $\varphi $. From (\ref{ddw}) we see the derivative do fail to commute at
the singular points of the unit vector $n^a$. The integral of the solid
angle density give the monopole charges 
\begin{eqnarray}
Q_m &=&\frac 1{4\pi }\int_\Sigma d\Omega   \nonumber \\
&=&\int_V\delta ^3(\varphi )J(\frac \varphi x)d^3x  \nonumber \\
&=&\int_{\varphi (V)}\delta ^3(\varphi )d^3\varphi   \nonumber \\
&=&\deg \varphi .
\end{eqnarray}
From the $\delta $-function theory we know that if $\varphi (x)$ has $%
l_\varphi $ isolated zeros and let the $i$th zero be $z_i$, $\delta (\varphi
)$ can be expressed as\cite{sch} 
\begin{equation}
\delta (\varphi )=\sum_{i=1}^{l_\varphi }\frac{\beta _i(\varphi )}{J(\frac %
\varphi x)}\delta (x-z_i).
\end{equation}
Then one obtains 
\begin{equation}
\delta ^3(\varphi )J(\frac \varphi x)=\sum_{i=1}^{l_\varphi }\beta
_i(\varphi )\eta _i(\varphi )\delta ^3(x-z_i).
\end{equation}
where $\beta _i(\varphi )$ is the positive integer (the Hopf index of the $i$%
th zero) and $\eta _i(\varphi )$ the Brouwer degree\cite{dub,mil} 
\begin{equation}
\eta _i(\varphi )=signJ(\frac \varphi x)|_{x=z_i}=\pm 1.
\end{equation}
From above deduction the topological structure of the solid angle density
projection is obtained 
\begin{equation}
d^2\Omega =4\pi \sum_{i=1}^{l_\varphi }\beta _i(\varphi )\eta _i(\varphi
)\delta ^3(x-z_i)d^3x,
\end{equation}
and the monopole charge is 
\begin{equation}
Q_m=\sum_{i=1}^{l_\varphi }\beta _i(\varphi )\eta _i(\varphi ).
\end{equation}

If we denote the complex field $\phi $ as 
\begin{equation}
\phi =||\phi ||e^{i\theta }  \label{cphi}
\end{equation}
where 
\begin{equation}
\tan \theta =\frac \sigma \rho .
\end{equation}
Then we get from (\ref{a}) and (\ref{cphi}) 
\begin{equation}
A=d\theta -\frac 1{2i||\phi ||^2}(D\phi \phi ^{*}-\phi D\phi ^{*}).
\end{equation}
For the topological charges is independent of choice of gauge, here we take
the complex field $\phi $ as a gauge constant 
\begin{equation}
D\phi =0.
\end{equation}
It is easy to prove under this gauge condition $A$ is the angle density 
\begin{equation}
A=d\theta .
\end{equation}
Noticing the singularity again we find the gauge field corresponding to $A$
is not exact at the singular points. Analogous to the deduction of the local
topological structure of monopole density and using the relationship 
\[
\frac{\phi ^A}{||\phi ||^2}=\frac \partial {\partial \phi ^A}\ln ||\phi
||\quad \quad A=1,2
\]
in which $\phi ^1=\rho $ and $\phi ^2=\sigma ,$ we find 
\begin{eqnarray}
d^2\theta  &=&\varepsilon ^{AB}d\frac{\phi ^A}{||\phi ||^2}\wedge d\phi ^B 
\nonumber \\
&=&\varepsilon ^{AB}\frac \partial {\partial \phi ^A}\frac \partial {%
\partial \phi ^C}\ln (||\phi ||)d\phi ^C\wedge d\phi ^B.
\end{eqnarray}
Defining Jacobian vector $J^\mu (\frac \phi x)$ 
\begin{equation}
\varepsilon ^{AB}J^\mu (\frac \phi x)=\varepsilon ^{\mu \nu \lambda
}\partial _\nu \phi ^A\partial _\lambda \phi ^B
\end{equation}
and using 2-dimensional Laplacian relation 
\begin{equation}
\frac \partial {\partial \phi ^a}\frac \partial {\partial \phi ^a}\ln ||\phi
||=2\pi \delta ^2(\phi ),
\end{equation}
we get 
\begin{equation}
d^2\theta =2\pi \delta ^2(\phi )J^\mu (\frac \phi x)d\sigma _\mu 
\end{equation}
where $d\sigma _\mu $ is area element of the surface 
\begin{equation}
d\sigma _\mu =\frac 12\varepsilon _{\mu \nu \lambda }dx^\nu \wedge
dx^\lambda .
\end{equation}
The derivative fails to commute again at the zero points of the complex
field $\phi $ and these points behave as the sources of the vortices. 
\begin{eqnarray}
Q_v &=&\frac 1{2\pi }\int_\Sigma d^2\theta   \nonumber \\
&=&\int_\Sigma \delta ^2(\phi )J^\mu (\frac \phi x)d\sigma _\mu   \nonumber
\\
&=&\int_\Sigma \delta ^2(\phi )d^2\phi   \nonumber \\
&=&\deg \phi .
\end{eqnarray}
Let us choose coordinates $y=(u^1,u^2,v)$ such that $u=(u^1,u^2)$ are
intrinsic coordinate on $\sum $. Suppose $\phi (x)$ possess $l_v$ isolated
zeros and denote the $i$th zero by $w_i$ and using the $\delta -$function
theory we get 
\begin{equation}
\delta ^2(\phi )J^\mu (\frac \phi x)=\sum_{i=1}^{l_v}\beta _i(\phi )\eta
_i(\phi )\delta ^2(u-w_i)\frac{dy_i^\mu }{dv}.
\end{equation}
Then the local structure of the vortex density is 
\begin{equation}
f=2\pi \sum_{i=1}^{l_v}\beta _i(\phi )\eta _i(\phi )\delta ^2(u-w_i)d^2u
\end{equation}
and the vorticity is 
\begin{equation}
Q_v=\sum_{i=1}^{l_v}\beta _i(\phi )\eta _i(\phi ).
\end{equation}
Therefore we obtain the total topological charge 
\begin{eqnarray}
Q &=&Q_m-\frac 12Q_v  \nonumber \\
&=&\sum_{i=1}^{l_m}\beta _i(\varphi )\eta _i(\varphi )-\frac 12%
\sum_{i=1}^{l_v}\beta _i(\phi )\eta _i(\phi )  \nonumber \\
&=&\deg \varphi -\frac 12\deg \phi ,
\end{eqnarray}
which is fractional quantized in terms of $\frac 12$. The corresponding
disclination 
\begin{eqnarray}
S &=&4\pi Q_m-2\pi Q_v  \nonumber \\
&=&2\pi (2Q_m-Q_v)
\end{eqnarray}
is quantized with $2\pi $ as the unit disclination charge.

\section{Algebra in Chern-Simons gravity}

In Chern-Simons theory of the three dimensional gravity there exists an
affine Kac-Moody algebra (at the boundary). The canonical generator $Q(n)$
associated to a gauge transformation, $\delta A^a=Dn^a=[A^a,Q(n)]$, any
three-dimensional Chern-Simons theory is given by\cite{bal,ba4} 
\begin{equation}
Q(n)=\frac k{4\pi }\int_\Sigma n^aF^a-\frac k{4\pi }\int_{\partial \Sigma
}n^aA^a,  \label{qn}
\end{equation}
where $\Sigma $ is a two-dimensional spatial section with boundary $\partial
\Sigma $, $F$ is the 2-form curvature and $A$ is the gauge potential. It is
easy to check that the boundary term arising when varying the bulk part of (%
\ref{qn}) is cancelled by the boundary term. The Poisson bracket of two
Functions $F(A_i)$ and $H(A_i)$ is computed as 
\begin{equation}
\{F,H\}=\frac{4\pi }k\int_\Sigma d^2z\frac{\delta F}{\delta A_i^a(z)}%
\varepsilon _{ij}\frac{\delta H}{\delta A_j^a(z)}.  \label{p-b}
\end{equation}
By direct application of the Poisson bracket (\ref{p-b}) one can find the
algebra of two transformations with parameters $n$ and $m$%
\begin{equation}
\lbrack Q(n),Q(m)]=Q([n,m])+{\frac k{4\pi }}\int_{{\partial }\Sigma }n^adm^a,
\label{a6}
\end{equation}
where $[n,m]=\varepsilon ^{abc}n^bm^c$. When $n=0$ at the boundary on can
find $Q(n)=G(n)$. Then there exist 
\begin{eqnarray}
\lbrack G,G] &=&G;  \label{a7} \\
\lbrack G,Q] &=&G.  \label{a8}
\end{eqnarray}
Equations (\ref{a6}) (\ref{a7})\ and (\ref{a8}) form a Kac-Moody algebra$%
\cite{brown,bal,ba4}$ with central charge is 
\begin{equation}
{\frac k{4\pi }}\int_{{\partial }\Sigma }n^adm^a.
\end{equation}

For black hole problem the choice of gauge group in Euclidean signature is $%
SO(3)$. From (\ref{qn}) we see the generator $Q(n)$ just the topological
charge amended for the manifold with a boundary 
\begin{eqnarray}
Q(n) &=&\frac k{4\pi }\int_\Sigma (dA-d\Omega )-\frac k{4\pi }\int A 
\nonumber \\
&=&-\frac k{4\pi }\int_\Sigma d\Omega   \nonumber \\
&=&-kQ_m(n)
\end{eqnarray}
which means the generator $Q(n)$ is the solid angle of the surface $\Sigma $
corresponding the singular points of $n^a.$ One should noticed here the
monopole charge $Q_m$ need not to be an integral for the surface $\Sigma $
is not closed. Now the commutative relation (\ref{a6}) is rewritten as 
\begin{equation}
\lbrack kQ_m(n),kQ_m(m)]=-kQ_m[n.m]+\frac k{4\pi }\int_{\partial \Sigma
}n^adm^a.
\end{equation}

After the gauge is fixed the value of $Q$ given in (\ref{qn}) reduces to the
boundary term 
\begin{equation}
\hat{Q}(n)=-\frac k{4\pi }\int_{\partial \Sigma }n^aA^a.
\end{equation}
A theorem\cite{brown} states that after the gauge is fixed and one works
with the induced Poisson bracket (or Dirac bracket), the charge $\hat{Q}$
satisfies the same algebra (\ref{a6}) as it did the full charge $Q$%
\begin{equation}
\lbrack \hat{Q}(n),\hat{Q}(m)]=\hat{Q}([n,m])+{\frac k{4\pi }}\int_{{%
\partial }\Sigma }n^adm^a.  \label{dbn}
\end{equation}
From (\ref{dbn}), we know the boundary term is just the vortex term of the
topological charge (\ref{t-c}) which is deduced as 
\begin{eqnarray}
\hat{Q}(n) &=&-\frac k{4\pi }\int_{\partial \Sigma }d\theta (n)  \nonumber \\
&=&-\frac k2Q_v(n).
\end{eqnarray}
Now we get the commutative relations of the vorticities 
\begin{equation}
\lbrack \frac k2Q_v(n),\frac k2Q_v(m)]=-\frac k2Q_v([n,m])+{\frac k{4\pi }}%
\int_{{\partial }\Sigma }n^adm^a.
\end{equation}

\section{Conclusion}

We have shown in this contribution that there exist two kinds of topological
structures in 3-dimensional Euclidean gravity by making use of the
decomposition of $SO(3)$ spin connection, namely the monopole structure and
vortex structure. When projection the disclination density onto a covariant
constant unit vector, the size of disclination is quantized topologically.
The topological charge equals to the sum of degrees of vectors $\varphi $
and $\phi $ which comes from the multiplet $(A_\mu ,\phi )$ in the
decomposition of gauge potential. The Hopf indices and Brouwer degrees of $%
\varphi $ and $\phi $ label the local structure of the topological defect.
Due to the singularity of the gauge potential, the topological densities of
disclination are found to be $\delta $-function of $\varphi $ and $\phi $.
The noncommutative properties of derivatives at the singular points reveal
the sources of the topological defect. We also showed the monopole charges
and vorticities can serve as the generators of Kac-Moody algebra in
3-dimensional Chern-Simons gravity.

\end{document}